\documentclass{PoS}

\usepackage{amsmath,amssymb}
\pdfoutput=1

\newcommand{\f}[2]{\frac{#1}{#2}}
\newcommand{\la}{\langle}
\newcommand{\ra}{\rangle}
\newcommand{\lla}{\la\!\la}
\newcommand{\rra}{\ra\!\ra}
\newcommand{\de}{\partial}

\newcommand{\B}{{\cal B}}

\title{$2D$ and $3D$ Antiferromagnetic Ising Model with 
 "topological" term at $\theta=\pi$}

\ShortTitle{$2D$ and $3D$ Antiferromagnetic Ising Model with topological term at $\theta=\pi$}

\author{Vicente Azcoiti\\
        Departamento de F\'isica Te\'orica, Universidad de Zaragoza, Calle Pedro Cerbuna 12, E-50009 Zaragoza, Spain\\
        E-mail: \email{azcoiti@azcoiti.unizar.es}}

\author{\speaker{Gennaro Cortese}\\
        Instituto de F\'isica Te\'orica
    UAM/CSIC, \sl Cantoblanco, E-28049 Madrid, Universidad Autonoma de Madrid \& Universidad de Zaragoza, Spain\\
        E-mail: \email{cortese@unizar.es}}

\author{Eduardo Follana\\
        Departamento de F\'isica Te\'orica, Universidad de Zaragoza, Calle Pedro Cerbuna 12, E-50009 Zaragoza, Spain\\
        E-mail: \email{efollana@unizar.es}}

\author{Matteo Giordano\\
        Institute for Nuclear Research of the Hungarian Academy of Sciences (ATOMKI), \sl Bem t\'er 18/c, H-4026 Debrecen, Hungary\\
        E-mail: \email{giordano@atomki.mta.hu}}

\abstract{We study the two and three-dimensional Antiferromagnetic Ising Model
  with an imaginary magnetic field $i\theta$ at $\theta = \pi$. We use
  a new geometric algorithm which does not present a sign
  problem. This allows us to perform efficient numerical simulations
  of this system.}

\FullConference{31st International Symposium on Lattice Field Theory - LATTICE 2013\\
		July 29 - August 3, 2013\\
		Mainz, Germany}

\begin{document}

\section{Introduction}

Numerical simulations are one of the main tools that allow us to obtain quantitative results 
in many fields of physics, and they have played an essential part in the 
progress of QCD and condensed matter physics in recent years. 
There are, however, many interesting physical systems for which 
we do not have efficient numerical algorithms yet. Examples include QCD at finite density or with a
non-vanishing $\theta$ term. This situation has hindered progress in
such fields for a long time, and it is thus of great interest to study
novel simulation algorithms.

In the present work we develop and test a geometric algorithm which we apply to study 
the two-dimensional antiferromagnetic Ising model with an
imaginary magnetic field i$\theta$ (see~\cite{primo},~\cite{secondo} and~\cite{Wu}) at $\theta = \pi$~\cite{quarto}, 
and which solves the sign problem that this model has when using standard algorithms. In this work also the $3D$ model is studied.

\section{The model and a geometric algorithm}

We start with the (reduced) Hamiltonian for the Ising model in $D\geq 2$ in an
external magnetic field,

\begin{equation}
  \label{eq:1}
  H(\{s_x\},F,h) = -F \sum_{(x,y)\in \B} s_x s_y -\frac{h}{2}\sum_{x} s_x\,.
\end{equation}
We consider a $D$-dimensional hypercubic lattice with 
$L=2n\in \mathbb{N}$ sites by side, and we denote the lattice sites by $x$ 
and the spin variables with $s_x \in \{\pm 1\}$. The sum $\sum_{(x,y)\in
  \B}$ runs over the set $\B$ of all nearest-neighbors $(x,y)$; $F =
J/(k_{B}T)$ is the reduced coupling between spins, and $h = 2B/(k_{B}T)$,
with $B$ the external magnetic field. The total number of spins is
$V=L^D=(2n)^D$, thus even, and therefore the quantity $Q\equiv\f{1}{2}\sum_{x} s_x$ is an
integer number, taking values between $-L^D/2$ and $L^D/2$, which can
be thought of as playing the role of a topological charge. We are
interested in studying this system for imaginary values of the reduced
magnetic field $h$, i.e., for $h=i\theta$.

This model suffers from a sign problem, because the weight of a
configuration is not a positive real number, and therefore we cannot
apply a standard algorithm. For the special case $\theta = \pi$ we
will show how to construct an efficient geometric algorithm that
circumvents this problem.

The partition function of the system at $\theta = \pi$ is

\begin{equation}
  \label{eq:4}
  \begin{aligned}
  Z(F,\theta=\pi) = \sum_{\{s_x=\pm 1\}} e^{F\sum_{(x,y)\in \B} s_x
    s_y + i\f{\pi}{2}\sum_{z} s_z}  \\
= i^{V}\sum_{\{s_x=\pm 1\}} \left\{\prod_{(x,y)\in \B} [\cosh(F s_x s_y) +
\sinh(F s_x s_y)] \prod_z {s_z}\right\} \\
= \sum_{\{s_x=\pm 1\}} \left\{\prod_{(x,y)\in \B} [\cosh(F) +
\sinh(F)s_x s_y] \prod_z {s_z}\right\}
\,.
  \end{aligned}
\end{equation}
It is easy to see, by decomposing the lattice in two staggered
sublattices, that $Z(F, \pi) = Z(-F, \pi)$; therefore, at $\theta =
\pi$, the ferromagnetic and antiferromagnetic models are equivalent, 
and furthermore $Z$ is in fact a function of
$|F|$.

We can now expand the product inside the curly brackets, and assign a
unique graph to each term in the expansion: to each factor $\sinh(|F|)
s_x s_y$ we assign the bond $(x,y) \in \B$ in the graph, and we say
that such a bond is active. Every other bond is called inactive, and
corresponds to a factor $\cosh(|F|)$. Each set of active bonds
describes one and only one of the terms in the expansion. After
summing over spin states most of the contributions in the expansion
vanish; only the graphs such that every node on the lattice has an odd
number of active bonds touching it give a non-vanishing
contribution. We call such graphs admissible, and we denote by ${\cal
  G}$ the set of all admissible graphs. If we consider an admissible
graph, $g \in {\cal G}$, and denote by ${\cal N}_b(g)$ the number of
active bonds in $g$, by $\bar{\cal N}_b(g)$ the number of inactive
bonds in $g$, and by ${\cal N}$ the total number of bonds in the
lattice (thus ${\cal N} = {\cal N}_b(g) + \bar{\cal N}_b(g)$), the
weight of such graph in the partition function is
$2^V\sinh(|F|)^{{\cal N}_b(g)}\cosh(|F|)^{\bar{\cal N}_b(g)}$.
Therefore we can rewrite the partition function as:
\begin{equation}
  \label{eq:7opties}
  \begin{aligned}
    Z(F,\theta=\pi)&=  2^{V}\sum_{\substack{g\in {\cal G}}}
        \cosh(|F|)^{{\bar{\cal N}}_b(g)}
    \sinh(|F|)^{{\cal N}_b(g)} \\ 
    &= 2^{V}\cosh(|F|)^{{\cal N}}\sum_{\substack{g \in
        {\cal G}}}
        \tanh(|F|)^{{\cal N}_b(g)}\,.
  \end{aligned}
\end{equation}
The important point here is that all configurations (graphs) have
positive weights, and therefore this representation of the partition
function does not have a sign problem.
Now, generalizing the partition function~(\ref{eq:7opties}) to the case where 
the coupling is position-dependent, we obtain: 
\begin{equation}
  \label{eq:9}
  Z(\{F_{xy}\},\theta=\pi) =\sum_{\{s_x=\pm 1\}} \left\{\prod_{(x,y)\in \B}
  [\cosh(|F_{xy}|) + 
  \sinh(|F_{xy}|)s_x s_y] \prod_z {s_z} \right\}\,.
\end{equation}
From this expression we are able to build all the correlation functions for an
even number of spins (correlation functions with an odd number 
of spins are automatically zero).
After some calculations (see~\cite{paper}) we obtain the following expression for the correlation functions:
 \begin{multline}
  \label{eq:14bis}
  C(d,F)\equiv \la s_x s_{x+d\hat 1} \ra = \left[\prod_{i=1}^d \f{\de}{\de F_{x_i
      x_{i+1}}}\right] \log
Z(\{F_{xy}\},\theta=\pi)\bigg|_{\{F_{xy}\}=\{F\}}\\ =  \lla
  \tanh(F)^{d-2{\cal N}_b[g,\{x_{i},y_{i}\}]} \rra\,,
\end{multline}
where ${\cal N}_b[g,\{x_{i},y_{i}\}]$ is the number of active bonds 
along the straight path connecting $x$ and
$x+d\hat 1$. However, we stress the fact that the specific choice of
the path is irrelevant, as they are all equivalent, as long as the
endpoints are fixed. It is also possible to obtain the relations for the energy 
density and specific heat by taking derivatives of the partition function with respect to the 
coupling constant. In fact, using our geometric algorithm these observables are 
related to the average number of active bonds and its fluctuations (the details of 
the derivation of these expressions will be discussed in~\cite{paper}). 

\section{Simulation}

In order to perform calculations using Monte Carlo methods, we
need an efficient algorithm to explore the space of configurations,
that in the geometric representation is given by the set of admissible
graphs, ${\cal G}$. The essential ingredient is a local
prescription\footnote{Local here means that only a fixed number of
  bonds in a bounded region are changed when updating a configuration.} 
  that takes the system from
an admissible configuration to another admissible configuration. The
simplest change that one can do to a configuration is to make an
inactive bond into an active one, or viceversa. However, applying this
change to an admissible configuration will not take us to another
admissible configuration. Let us consider instead an arbitrary square
on the lattice, and consider all possible changes in the state of the
bonds in the square. It is easy to convince oneself that if we start
from an admissible configuration, the only way to arrive at another
admissible configuration is either by not changing the state of any of
the bonds, or by changing all of them. Using these steps and the
weights of the configurations derived above, it is now easy to set up
a standard Metropolis algorithm\footnote{In order to achieve ergodicity 
for periodic boundary conditions we have to implement 
also a few global steps (more details in [5]).}.

\begin{figure}[ht]
\begin{center}
\includegraphics[scale=0.26]{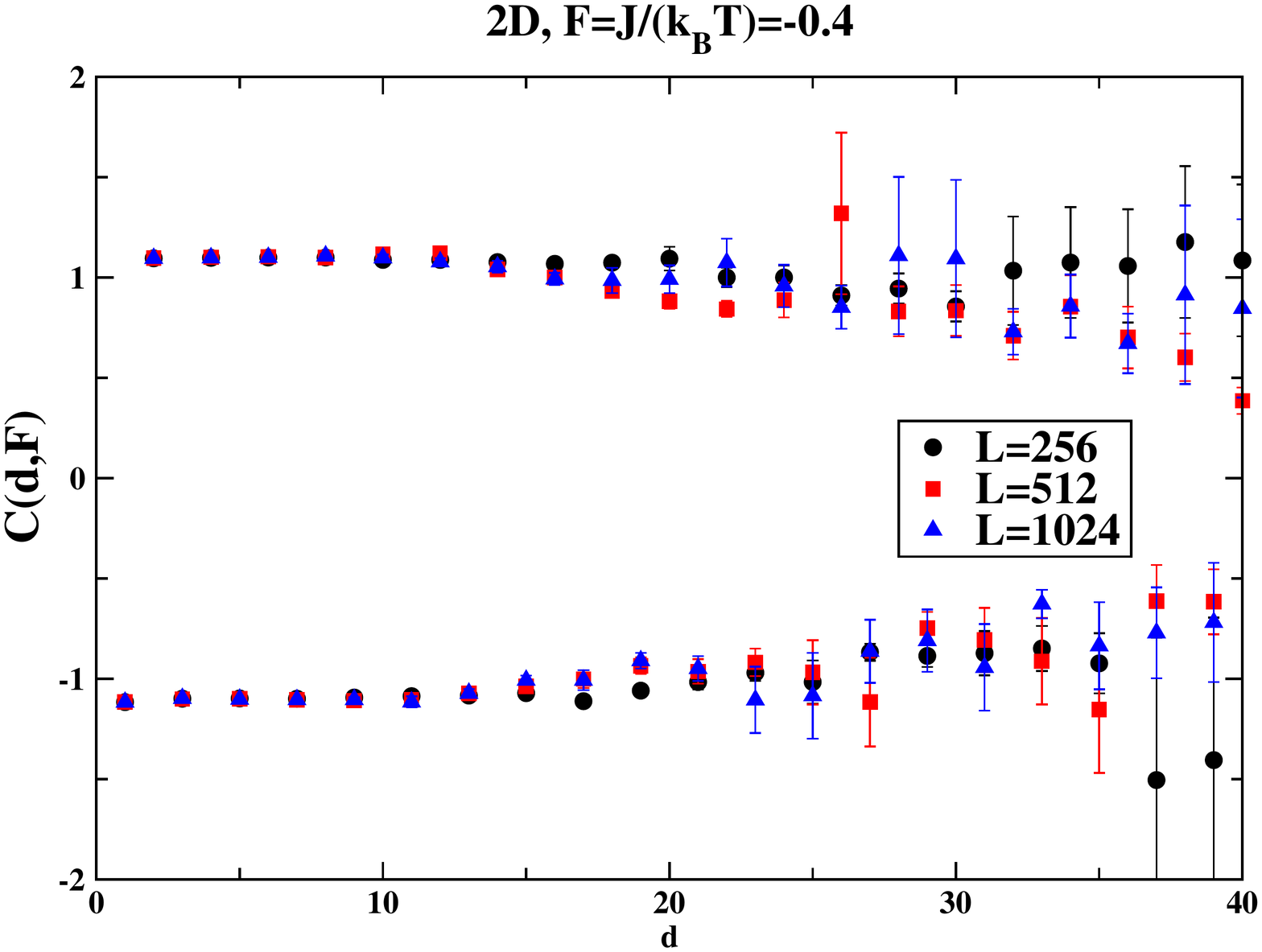}
\includegraphics[scale=0.26]{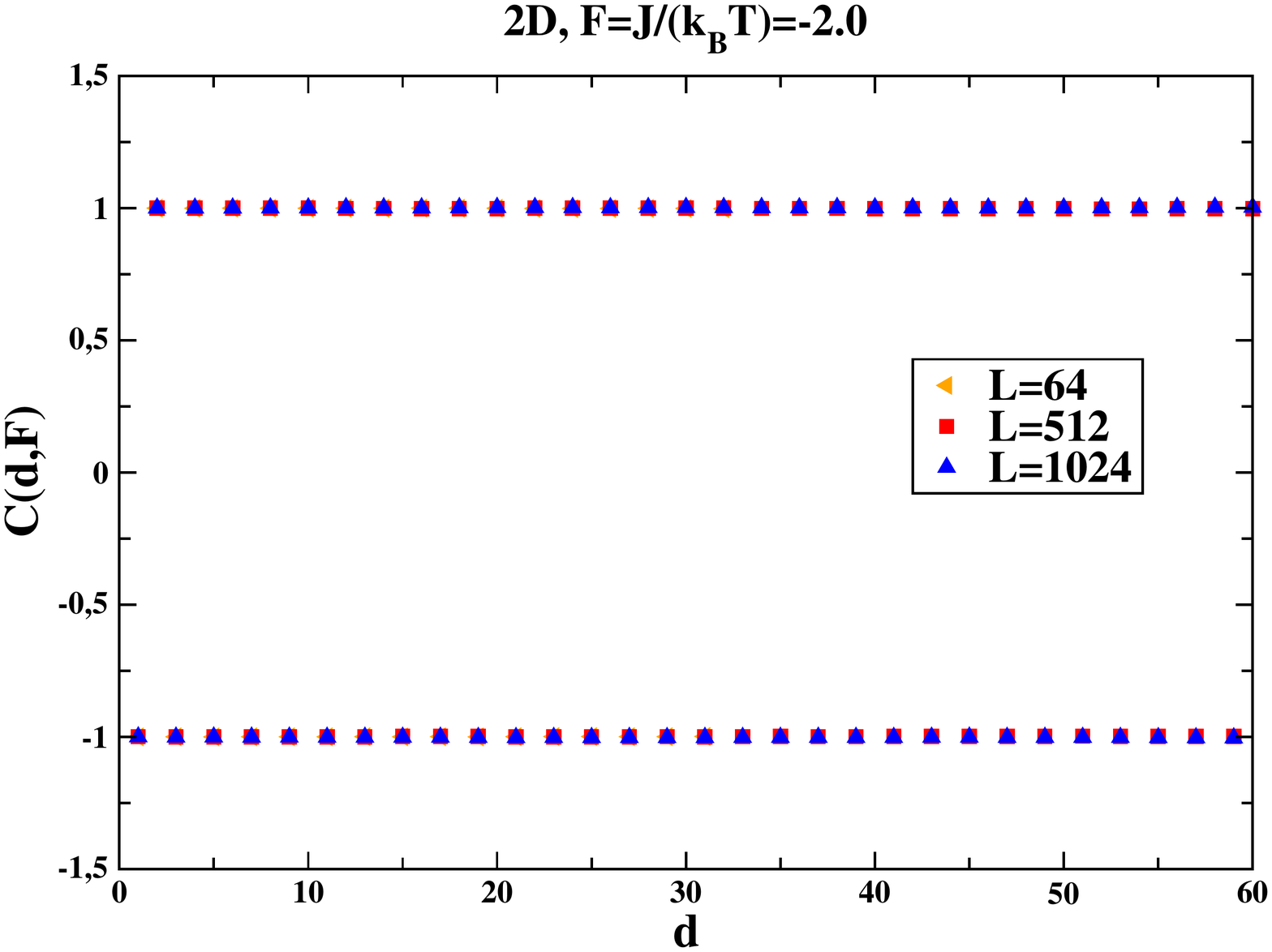}

\includegraphics[scale=0.26]{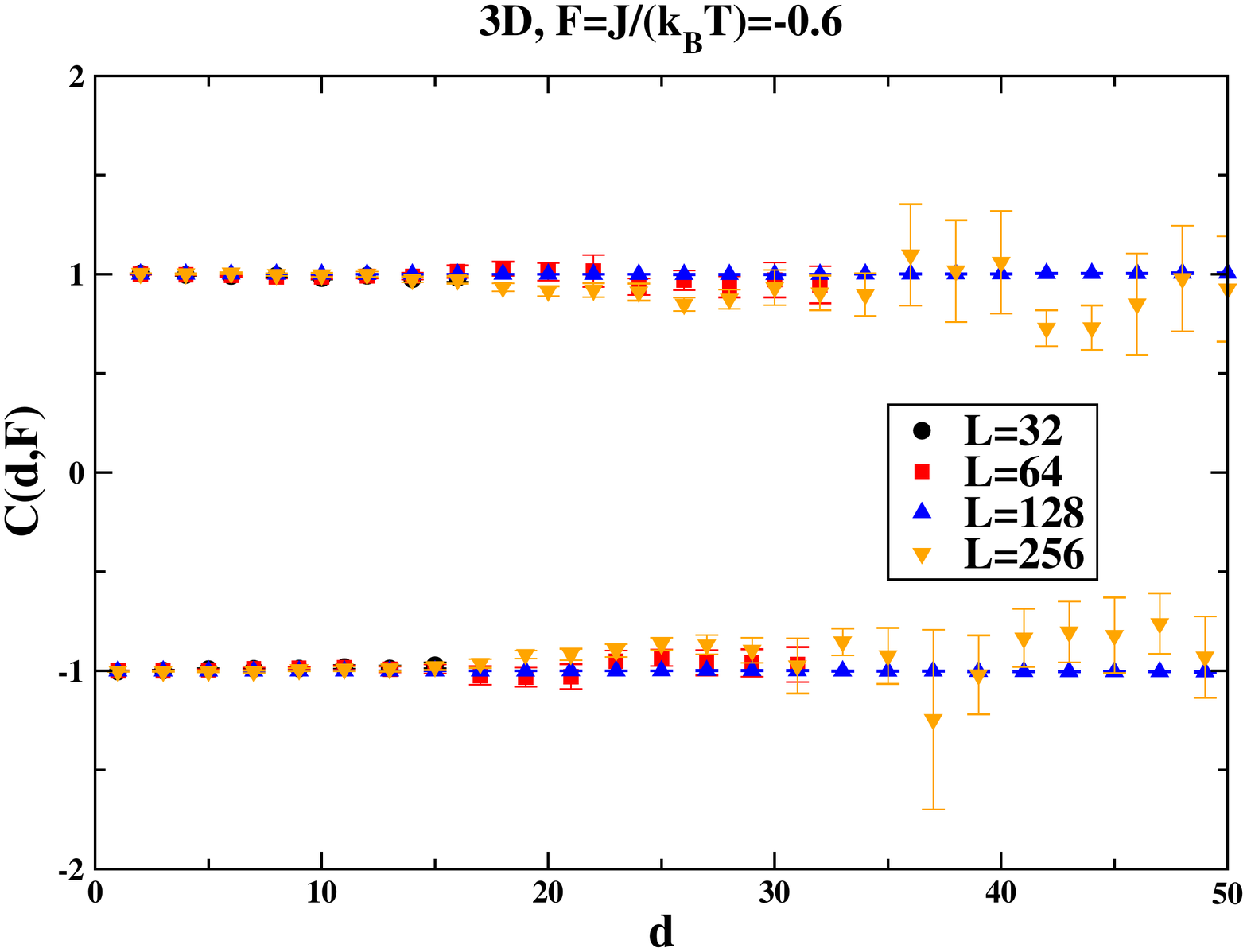}
\includegraphics[scale=0.26]{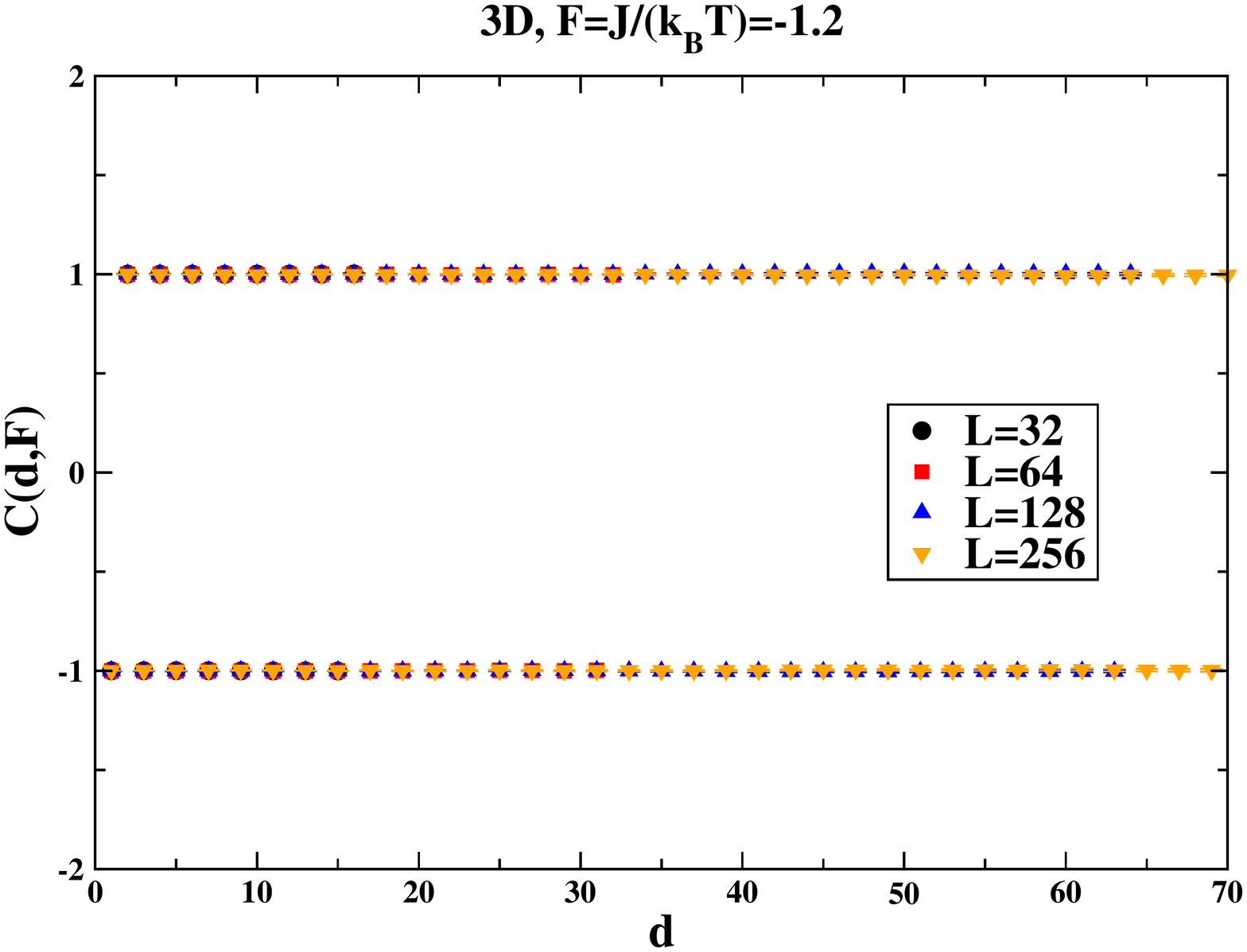}
\end{center}
\caption{Dependence of the correlation functions on the distance $d$ for some 
values of the coupling $F$ and for various lattice sizes $L$ for the $2D$ model (top) and 
the $3D$ model (bottom).} 
\label{corr1}
\end{figure}

We measured how the correlation functions~(\ref{eq:14bis}) depends on the distance $d$,
choosing different antiferromagnetic couplings $F<0$ and varying the lattice volume $V=L^D$, in order to determine the staggered magnetization from their asymptotic behavior.
In Figs.~(\ref{corr1}) we present some results concerning the evaluation of these 
correlation functions for different values of the coupling 
$F$ and various lattice sizes ranging from $L=64$ to $L=1024$ both for the $2D$ model and the $3D$ model. Simulations are done 
collecting 100k measurements for each value of $F$. For each run we discarded the first 10k
configurations in order to ensure thermalization. The jackknife method over bins at different blocking levels 
was used for the data analysis. 
%%%%%%%%%%%%%%%%%%%%%%%%%
%%%%%%%%%%%%%%%%%%%%%%%%%

\begin{figure}[ht]
\begin{center}
\includegraphics[scale=0.26]{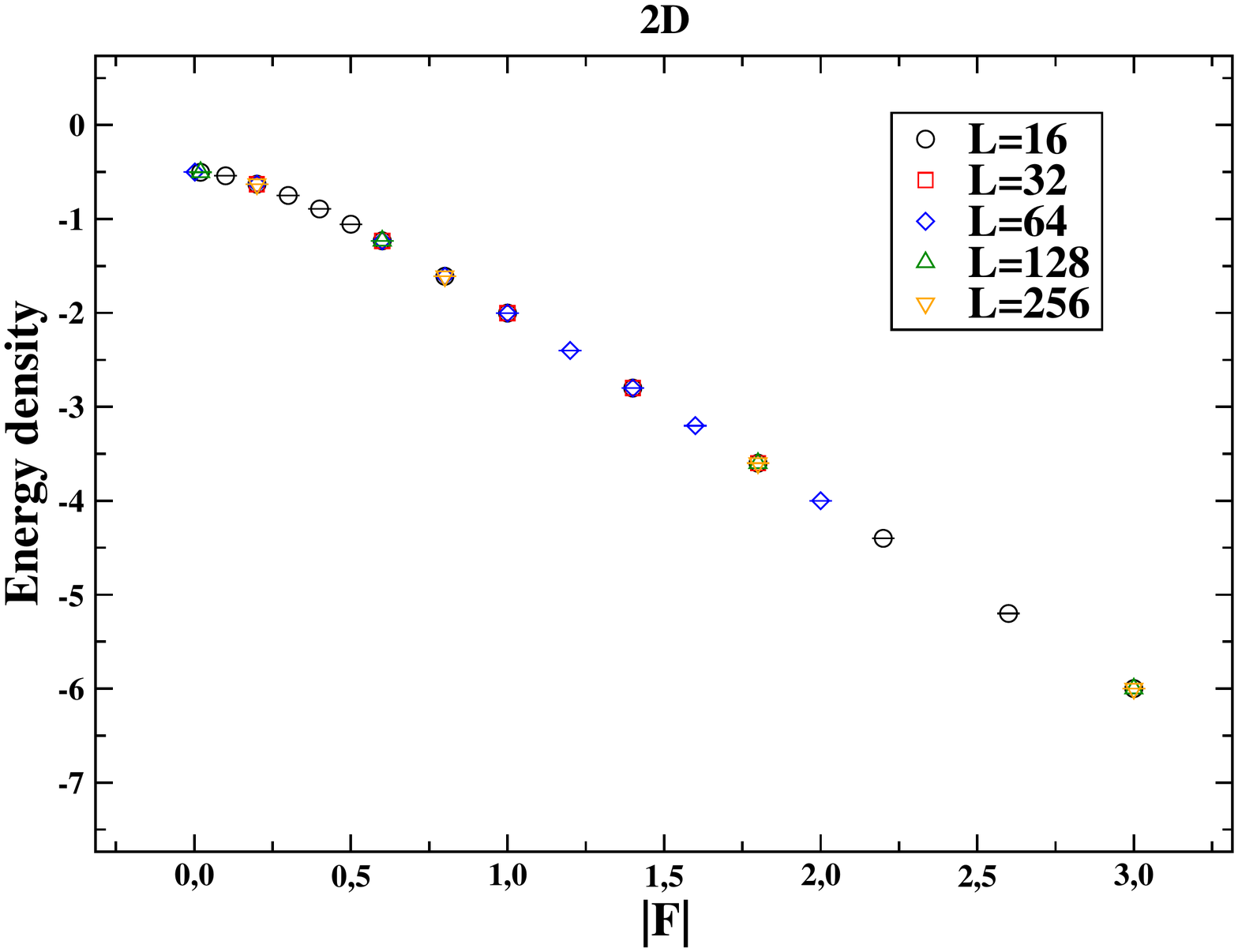}
\includegraphics[scale=0.26]{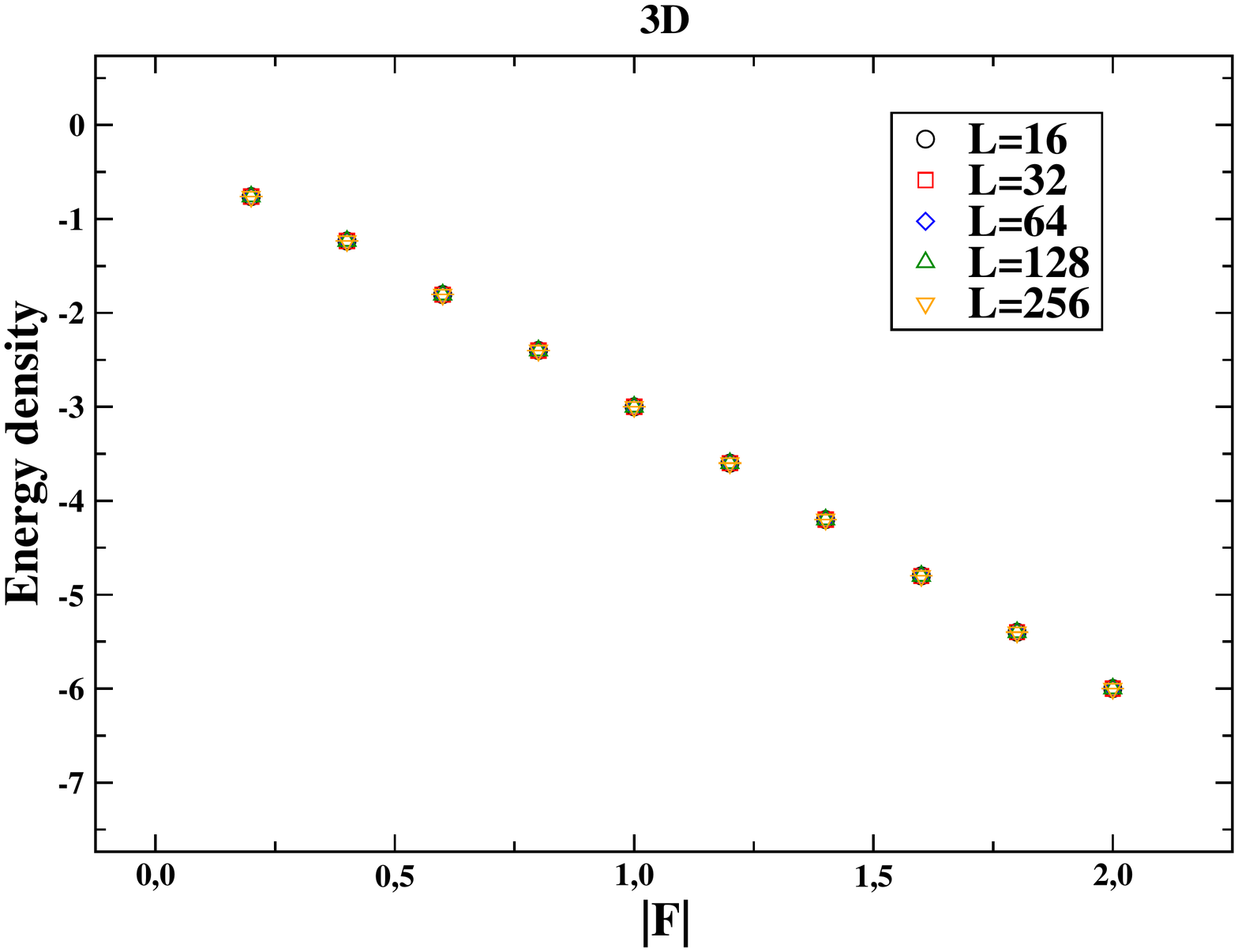}

\includegraphics[scale=0.26]{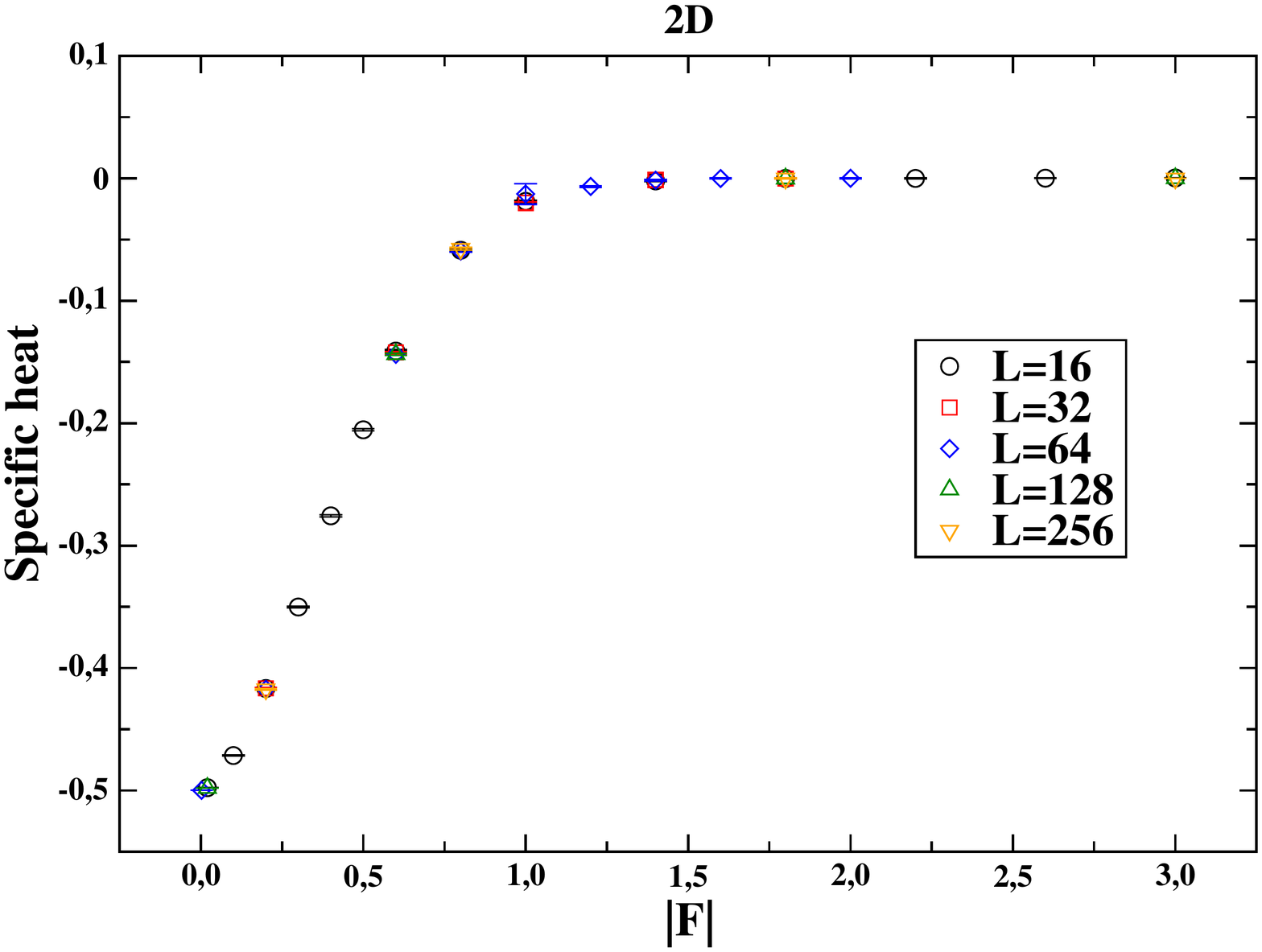}
\includegraphics[scale=0.26]{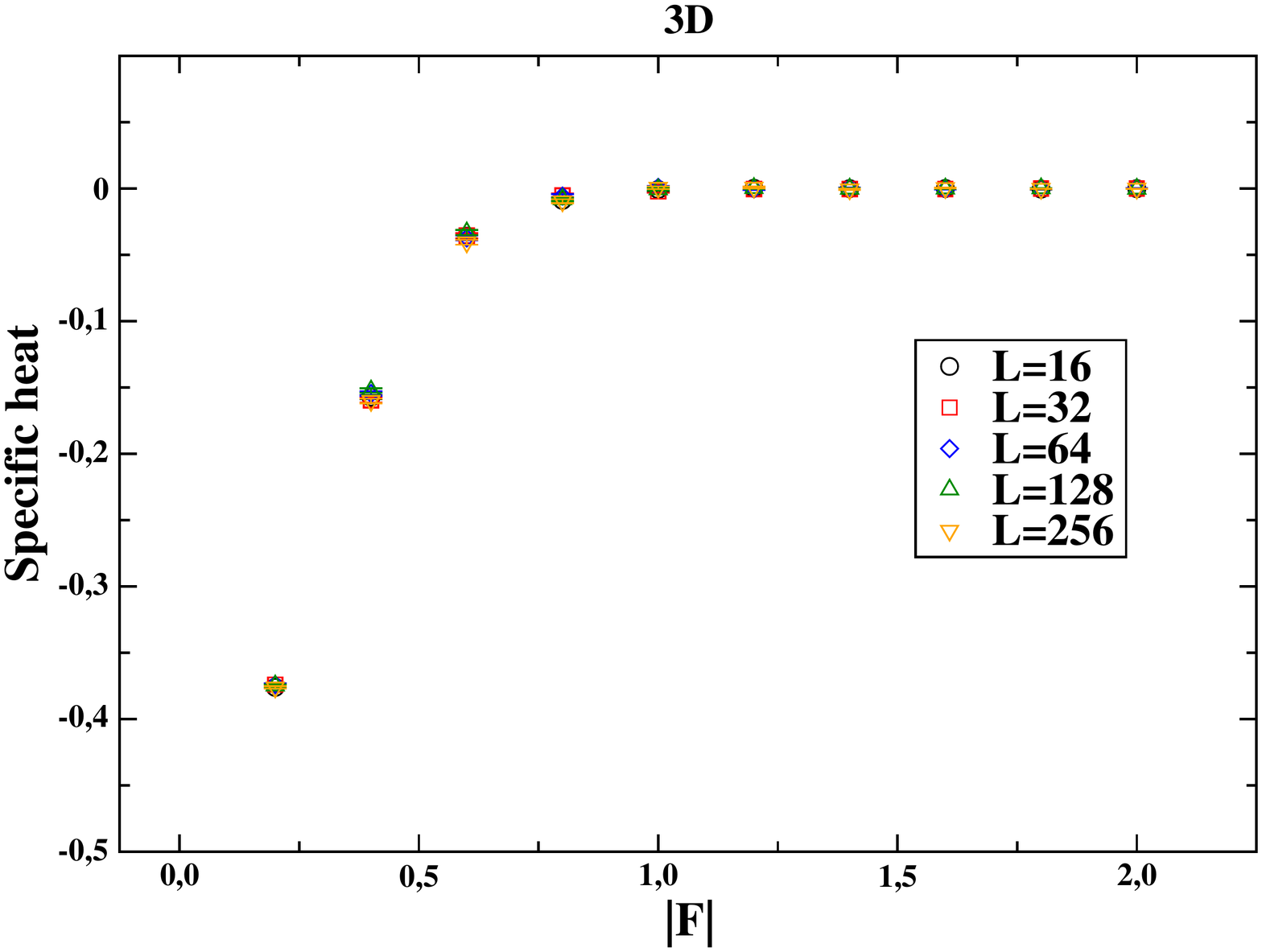}
\end{center}
\caption{Energy density as a function of $|F|$ for various lattice sizes $L$ 
for the $2D$ model (top left) and the $3D$ model (top right); specific heat as a function of $|F|$ for 
the $2D$ model (bottom left) and the $3D$ model (bottom right).} 
\label{energy}
\end{figure}

\begin{figure}[ht]
\begin{center}
\includegraphics[scale=0.35]{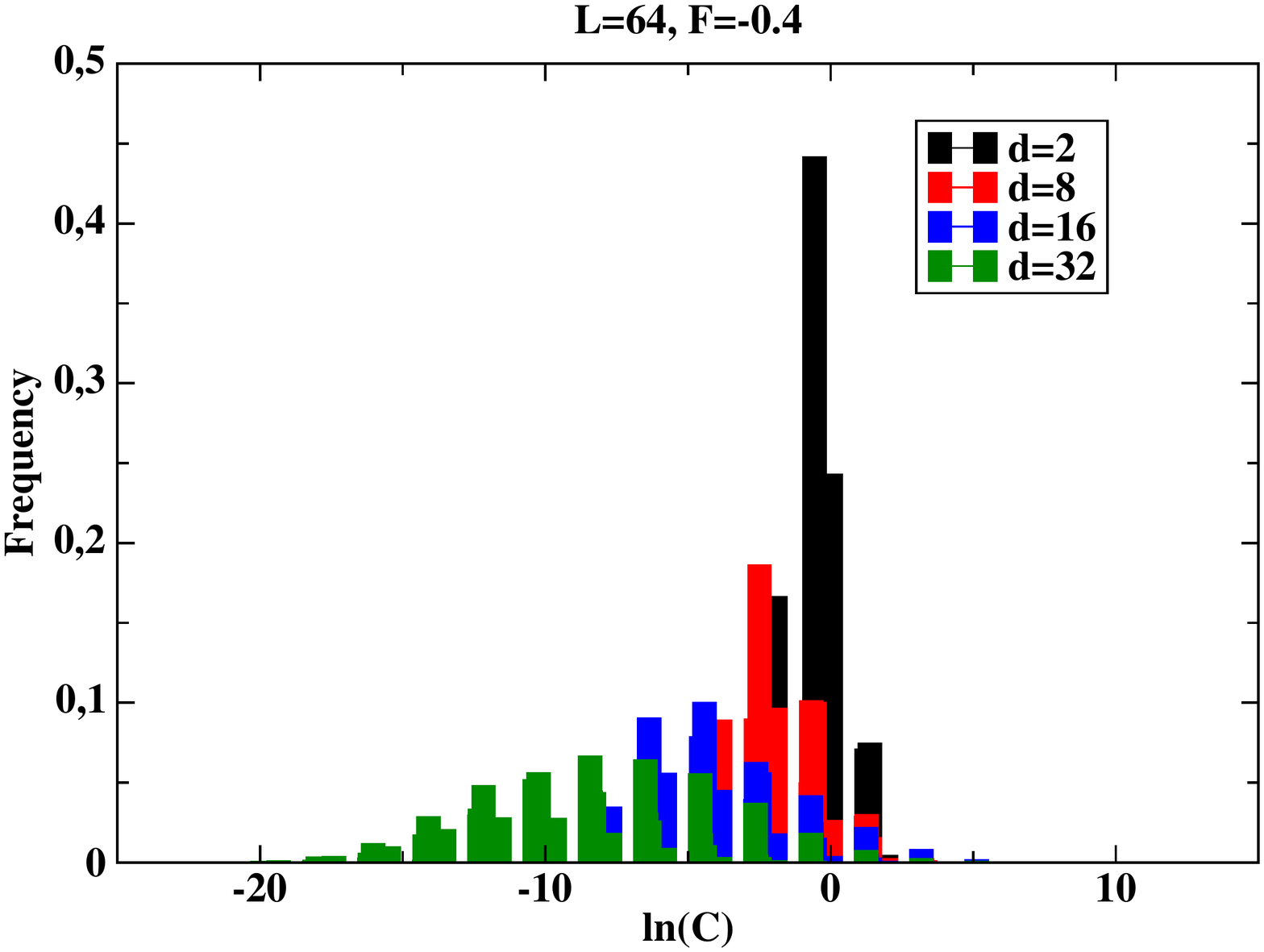}
\includegraphics[scale=0.35]{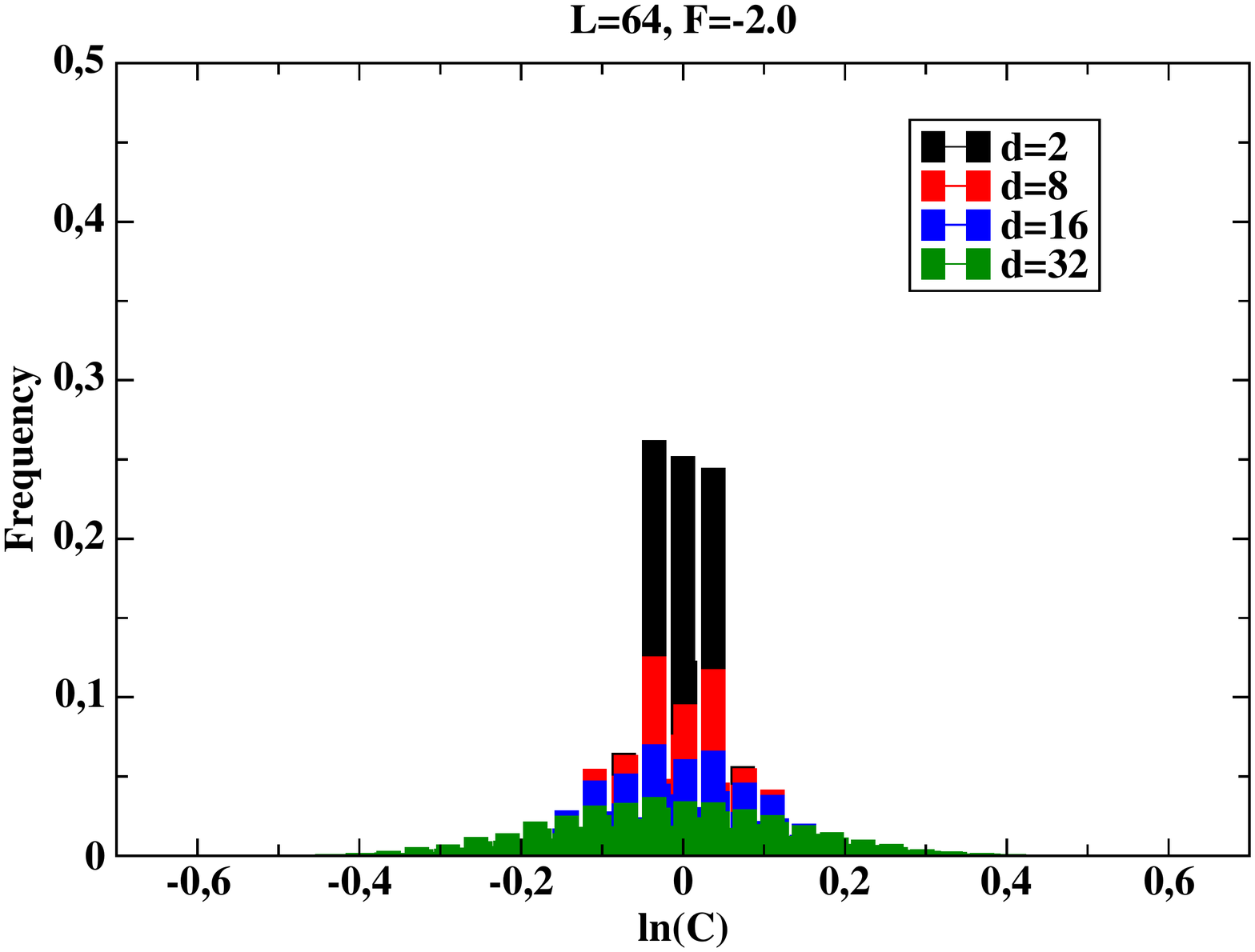}
\end{center}
\caption{Probability distribution of the logarithm of the correlator $C(d,F)$ for various distances $d$, for a lattice of size $L=64$, and 
for coupling $F=-0.4$ (top) and $F=-2.0$ (bottom). The figures depicted here are obtained for the $2D$ model.} 
\label{distribution_log}
\end{figure}

For antiferromagnetic couplings, the behavior of the correlation functions 
implies that the staggered magnetization is nonzero, while the
total magnetization vanishes, in the whole range of couplings that we
investigated. This result is in agreement with Refs.~\cite{primo,secondo,Wu}, and with the
mean-field calculation of Ref.~\cite{rif1}. The apparent decrease of $C(d,F)$,
starting from  $d\sim 10\div 20$, for small values of $|F|$, is probably
due to the heavy-tailed probability distributions of the correlators,
which is also the cause of their noisy behavior. In Fig.~\ref{distribution_log} we show the
probability distributions of the logarithm of the correlators for
lattice size $L=64$ and $F=-0.4$ and $F=-2.0$. We can clearly see that
for a small coupling $|F|$ the values are spread in a wider range than for
$F=-2.0$, and also that a long tail is developed for large
distances. This makes more difficult to obtain an accurate numerical evaluation of the correlators. 
In Fig.~(\ref{energy}) we show also the behavior of the energy density and of the specific heat for both 
the $2D$  model and the $3D$ model signalling that no phase transitions occur. 
More details, also concerning the proof of the ergodicity of the algorithm, will be presented in~\cite{paper}.

\acknowledgments
The work was funded by MICINN (under grant FPA2009-09638 and FPA2008-10732), DGIID-DGA (grant 2007-E24/2), and by the EU under ITN-STRONGnet (PITN-GA-2009-238353). EF is supported by the MICINN Ramon y Cajal program. MG is supported by the Hungarian Academy of Sciences 
under ``Lend\"ulet'' grant No. LP2011-011, and partially by MICINN under the CPAN
project CSD2007-00042 from the Consolider-Ingenio2010 program.

\end{document}